\documentclass[12pt]{iopart}

\usepackage[dvips]{graphicx}%
\usepackage{bm}
\usepackage{cite}
\usepackage{multirow}
\usepackage[normalem]{ulem} 
\usepackage[colorlinks=true,linkcolor=blue,citecolor=blue,urlcolor=blue]{hyperref}

\newcommand{\op}[1]{{\bm{#1}}}

\begin{document}

\title[Quantum Phase Diagrams of Matter-Field Hamiltonians]{Quantum Phase Diagrams of Matter-Field Hamiltonians II:
	Wigner Function Analysis}
\author{R. L\'opez-Pe\~na, S. Cordero, E. Nahmad-Achar, O. Casta\~nos}

\address{
Instituto de Ciencias Nucleares, Universidad Nacional Aut\'onoma de M\'exico, Apartado Postal 70-543, 04510 Cd. Mx., Mexico }

\ead{lopez@nucleares.unam.mx}

\date{\today}

\begin{abstract}
Non-classical states are of practical interest in quantum computing and quantum metrology. These states can be detected through their Wigner function negativity in some regions. In this paper, we calculate the ground state of the three-level generalised Dicke model for a single atom and determine the structure of its phase diagram using a fidelity criterion. We also calculate the Wigner function of the electromagnetic modes of the ground state through the corresponding reduced density matrix, and show in the phase diagram the regions where entanglement is present. A finer classification for the continuous phase transitions is obtained through the computation of the surface of maximum Bures distance.
\end{abstract}

%
%


\section{Introduction}

The Wigner function was introduced in 1932 to give a description of a quantum system in phase space~\cite{wigner32}. For some excellent reviews see~\cite{hillery84,lee95,petruccelli15}. Its description of a quantum system is complete in the sense that it allows for the calculation of all the quantities that the usual wave function gives, thus supplying all the information of the system in phase space. It is not a real distribution function in phase space because it can be negative, which precludes an ordinary probability interpretation. The initial importance of the Wigner function was that it allowed to treat quantum mechanics and thermodynamics on the same footing as their classical counterparts, making it easier to identify new effects in the quantum case. One of the first areas to adopt the Wigner function was that of optics, where it was employed to describe the coherence and the polarisation of optical fields~\cite{mandel65}, to explore the quantum effects in electron transport~\cite{barker83}, to investigate the transport in resonant tunneling devices~\cite{ravaioli85,jacoboni04,querlioz13}, to study wave propagation through media~\cite{mazar98}, to inquire into different theories of quantum dissipation~\cite{kohen97}, etc. 

The concept of entanglement, on the other hand, emerges with the Einstein-Podolsky-Rosen paradox~\cite{einstein35}, although it was Schr\"odinger who coined the term~\cite{schroedinger35}. In it, the nonlocal behaviour of the system is reflected in the correlation between distant points, which the Wigner function visually displays. When this occurs, negative values in the function appear as a consequence of the interference between distant regions in phase space~\cite{siyouri16}; for this reason the volume of the negative part of the Wigner function has been proposed as a measure of non-classicality of quantum states~\cite{kenfack04}.

Experimentally, the Wigner function for quantum optical systems can be reconstructed using homodyne tomography~\cite{smithey93}, field ionisation detectors~\cite{lutterbach97,bertet02}, photon-counting~\cite{banaszek99}, two-window heterodyne measurements~\cite{lee99}, etc. For non-separable laser beams through a toroidal mirror see~\cite{mey14}. For other classes of systems it can also be determined: one can mention two Bell states and the five-qubit Greenberger-Horne-Zeilinger (GHZ) spin Schr\"odinger cat state~\cite{rundle17};  a single harmonically trapped atom~\cite{leibfried96}; an ensemble of helium atoms formed by partially coherent illumination of a double slit~\cite{kurtsiefer97}, etc. This has practical importance in quantum information processing, because the Wigner function provides more information about the quantum system than any other quantum approach~\cite{weinbub18}.

In this work, we analyse the behaviour in phase space of the two radiation modes of light across the finite phase diagram of the quantum ground state of the Hamiltonian described by the generalized Dicke model. A finer description of the phase diagram is proposed, by using a definition of the minimum fidelity surface (or maximum Bures distance surface).We evaluate the Wigner function of the radiation modes for different points in parameter space, to depict the behaviour of the system, validating the classification into stable-continuous, unstable-continuous, and discontinuous transitions, given in~\cite{cordero20}.  The numerical expressions given for the Wigner function enable us to calculate the expectation values of all the observables for the corresponding electromagnetic mode. Also, we observe that the Wigner function for both modes in the normal region show a unimodal quasi-distribution, while in the collective region at least one of the functions shows a bimodal quasi-distribution (its negativity showing the quantumness of the state); hence the different regions in the phase diagram may be characterized by it.  In addition, the linear entropy is calculated, which presents a discontinuity when a change of parity in the ground state occurs. It vanishes for large values of the control parameters: for these large values the linear entropy only detects where the photon mode dominating the composition of the ground state changes. 

In section~\ref{Dicke-model} we introduce the generalisation of the Dicke model to describe the system. In section~\ref{fid-qpt} we explain how we use the minimum fidelity surface (maximum Bures distance surface) to calculate the quantum phase diagram of the ground state. Section~\ref{Wigner-calc} calculates the Wigner function of the field modes, and in section~\ref{quantum-transitions} we show the results obtained for a single atom in the $\Lambda$-configuration. Section~\ref{linearentropy} presents the calculation of the linear entropy for different subsystems of the model, which complements the results obtained via the Wigner function and correlates with them very well both qualitatively and in a quantitative way. Finally in section~\ref{conclusions} we give some concluding remarks.

\section{Dicke Generalised Models}
\label{Dicke-model}

\begin{figure*}
	\begin{center}
		{\includegraphics[width=0.25\linewidth]{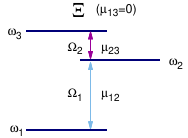}} \quad
		{\includegraphics[width=0.25\linewidth]{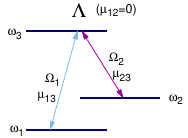}} \quad
		{\includegraphics[width=0.25\linewidth]{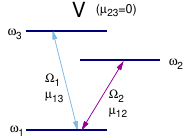}}
	\end{center}
        \caption{$3$-level atomic configurations. The $i$-th atomic energy level ($\hbar=1$) is denoted by $\omega_i$, and the coupling parameter between levels $i$ and $j$ is $\mu_{ij}$. The field frequencies are denoted by $\Omega_{ij}$. A particular atomic configuration is obtained by choosing appropriately the vanishing dipolar strength $\mu_{jk}$ in eq.(\ref{Hint}): for $\Xi$ we take $\mu_{13}=0$, for $\Lambda$ we take $\mu_{12}=0$, and for $V$ we take $\mu_{23}=0$. }
	\label{configurations}
   \end{figure*}

We consider the multipolar Hamiltonian for the dipole interaction between a two-mode radiation field and a $3$-level atomic system in the long wave approximation, which may be written as~\cite{castanos14, cordero15, nahmad-achar15} ($\hbar=1$)
   \begin{equation}
	\mathbf{H} = \mathbf{H}_{D} + \mathbf{H}_{int}\ ,
	\label{Hamiltonian3Levels}
   \end{equation}
where $\mathbf{H}_{D}$ is the diagonal matter and field independent contributions,
   \begin{equation}
      \mathbf{H}_{D}=\sum_{j<k}^{3} \Omega_{jk}\, \mathbf{a}_{jk}^\dag\, \mathbf{a}_{jk} + \sum_{j=1}^{3} \omega_j \, \mathbf{A}_{jj}\ ,
      \label{HD}
   \end{equation}
and $\mathbf{H}_{int}$ is the matter-field dipolar interaction
   \begin{equation}
      \mathbf{H}_{int}=-\frac{1}{\sqrt{N_a}} \sum_{j<k}^{3} \mu_{jk} \left(\mathbf{A}_{jk}+\mathbf{A}_{kj}\right)\left(\mathbf{a}_{jk} + \mathbf{a}_{jk}^\dag\right)\ .
      \label{Hint}
   \end{equation}
Here, $N_a$ denotes the number of particles, $\mathbf{a}_{jk}^\dag,\,\mathbf{a}_{jk}$ are the creation and annihilation photon operators for the mode $\Omega_{jk}$ which promotes transitions between the atomic levels $\omega_j$ and $\omega_k$, and $\mathbf{A}_{ij}$ are the matter operators obeying the $U(3)$ algebra
   \begin{equation}
       \left[\mathbf{A}_{ij},\,\mathbf{A}_{lm}\right] = \delta_{jl}\,\mathbf{A}_{im}-\delta_{im}\,\mathbf{A}_{lj}\ ,
   \end{equation}
with $\sum_{k=1}^3\mathbf{A}_{kk}=N_{a}\,\mathbf{I}_{\scriptsize\hbox{matter}}$. The coupling parameter between levels $\omega_j$ and $\omega_k$ has been denoted by $\mu_{ij}$, and we have assumed that the atomic frequencies satisfy $\omega_1 \leq \omega_2 \leq \omega_{3}$. We also fix $\omega_1=0$. Note that a particular atomic configuration is obtained by making an appropriate dipolar strength $\mu_{ij}$ to vanish (cf. Fig.~\ref{configurations}).

\begin{figure*}
	\begin{center}
		{\includegraphics[width=0.3\linewidth]{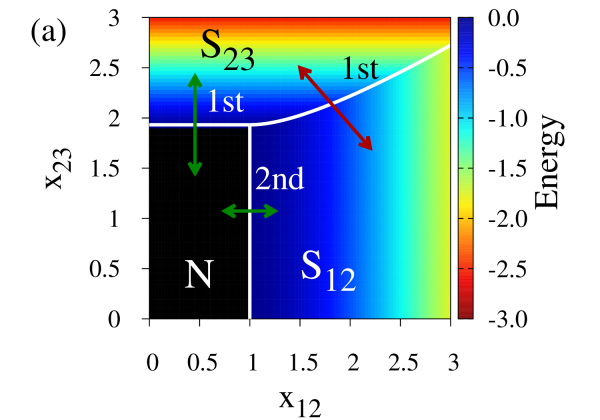}}
		{\includegraphics[width=0.3\linewidth]{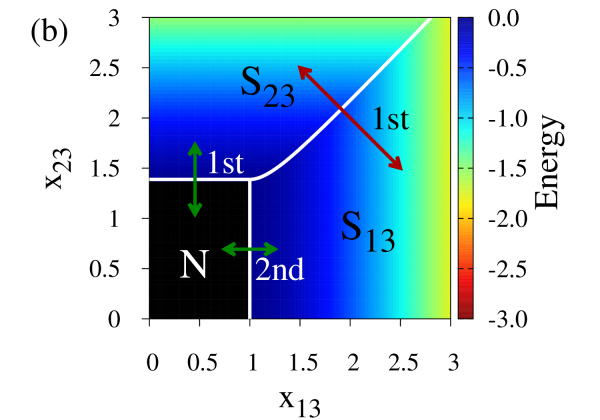}}
		{\includegraphics[width=0.3\linewidth]{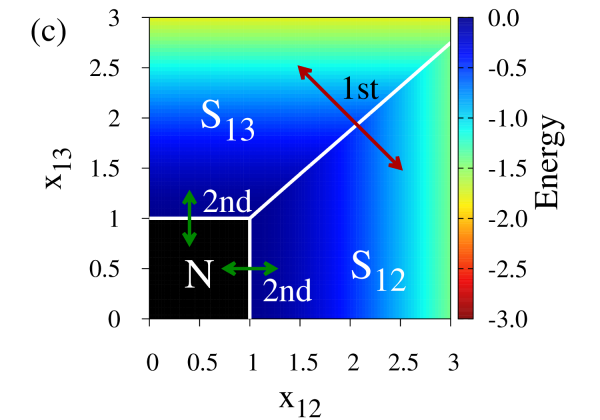}}
	\end{center}
        \caption{(colour online) Phase diagrams and energy surfaces, per particle, for the variational solution (thermodynamic limit) in the (a) $\Xi$-configuration with $\omega_2/\omega_3=1/3$, (b) $\Lambda$- configuration with $\omega_2/\omega_3=1/10$, and (c) $V$-configuration with $\omega_2/\omega_3=8/10$. The separatrices (white lines) and the order of the transitions are shown. The Normal regions (in black) are labeled by $N$. The superradiant regions divide themselves into subregions denoted by $S_{ij}$, where mode $\Omega_{ij}$ dominates. In all cases matter and field are in resonance, and the axes are $x_{ij} = \mu_{ij}/\mu^{c}_{ij}$, where $\mu^{c}_{ij}$ is the $2$-level critical coupling. In this and other plots, the energy is measured in units of $[\hbar \omega_3]$ and $x_{ij}$ is dimensionless.}
	\label{phasediagram}
   \end{figure*}

A variational study which involves coherent states for both matter and field contributions, provides a good approximation of the ground state energy surface per particle. The phase diagram in this approach shows the normal and collective regions, the latter divided into (two) regions where only one kind of photon contributes to the ground state, while the former remains in the vacuum state~\cite{cordero15}. This signature of the phase diagram remains when the symmetries of the Hamiltonian are restored in the variational solution and the thermodynamic limit is taken~\cite{cordero17}. In figure~\ref{phasediagram} the phase diagram and energy surface as functions of the dimensionless strengths $x_{jk}$ ($x_{jk}=\mu_{jk}/\mu_{jk}^c$ with $\mu_{jk}^c$ the critical value of the corresponding two level system) for the three atomic configuration are given, the separatrix (points where a sudden change in the ground state composition takes place) is draw in white lines, and the order of the transitions (using the Ehrenfest classification~\cite{gilmore93}) is shown. In the normal region (in black), labeled as $N$ and where atoms emit and absorb independently, both photons are in the vacuum state and the matter contribution is in its lowest energy state; while in the collective regions (gradient coloured regions), labeled as $S_{jk}$ and where superradiance takes place, only photons of kind $\Omega_{jk}$ and atomic populations in the levels $\omega_j$ and $\omega_k$ have non-zero contribution.

An exact calculation of the ground state involves a numerical diagonalisation of the Hamiltonian matrix. The Hamiltonian~(\ref{Hamiltonian3Levels}) is invariant under parity transformations of the form
   \begin{equation}
       \mathbf{\Pi}_{1}=e^{i \, \pi \, \mathbf{K}_1} \, , \quad \mathbf{\Pi}_2 = e^{i \, \pi \, \mathbf{K}_{2}}\ ,
       \label{parityoperators} 
   \end{equation}
where $\mathbf{K}_{s},\ s=1,2$, are constants of motion when the rotating wave approximation (RWA) is taken, which are found through the conditions $[\mathbf{\Pi_j},\,\mathbf{H}]=0$. Assuming that $\mathbf{K}_{s}$ is a linear operator, we find
   \begin{equation}
       \mathbf{K}_s = \eta^{(s)}_{12} \, {\bm \nu}_{12} +  \eta^{(s)}_{13} \, {\bm \nu}_{13} + \eta^{(s)}_{23} \, {\bm \nu}_{23} + \sum^3_{k=1}\lambda^{(s)}_k \, \mathbf{A}_{kk}\ .
       \label{eq.Kj3l}
   \end{equation}
where ${\bm \nu}_{12}$, ${\bm \nu}_{13}$, and ${\bm \nu}_{23}$ denote the number of photons of each electromagnetic field mode. The coefficients $\eta^{(s)}_{ij}$ and $\lambda^{(s)}_k$ of the operators are given in table~\ref{t.K3l} for the different atomic configurations. The operators shown $\mathbf{K}_s$ were chosen as linear combinations of constants of motion with non-negative integer eigenvalues.

\begin{table}
\caption{Coefficients $\eta^{(s)}_{ij}$ and $\lambda^{(s)}_k$ corresponding to the operators $\op{K}_s$ in eq.~(\ref{eq.Kj3l}) are given for the atomic $\Lambda$-, $\Xi$- and $V$-configurations.}
\label{t.K3l}
\begin{center}
\begin{tabular}{c c | c c c c c c}
Conf. &  $\op{K}_s$ & $\eta_{12}^{(s)}$ & $\eta_{13}^{(s)}$ & $\eta_{23}^{(s)}$ &$\lambda_{1}^{(s)}$ & $\lambda_{2}^{(s)}$ & $\lambda_{3}^{(s)}$\\ \hline\hline & & & & & & & \\[-3mm]
$\Lambda$&$\op{K}_1$ &  0 &  1 & 1 & 0 & 0 &1 \\[2mm]
&$\op{K}_2$ & 0 &0 &1 &1 &0 &1 \\[1mm] \hline\hline & & & & & & & \\[-3mm]
$\Xi$&$\op{K}_1$ & 1 & 0 & 1 & 0 & 1 & 2\\[2mm]
&$\op{K}_2$ & 0 & 0 & 1 & 0 &0 &1\\[1mm] \hline\hline & & & & & & & \\[-3mm]
$V$&$\op{K}_1$ & 1 & 1 & 0 & 0 & 1 &1 \\[2mm]
&$\op{K}_2$ & 0 & 1 & 0 & 0 & 0& 1 \\[1mm] \hline\hline 
\end{tabular}
\end{center}
\end{table}

Accordingly, the Hilbert space ${\mathcal H}$ divides naturally into four subspaces of the form 
   \[
      {\cal H}={\cal H}_{ee}\oplus{\cal H}_{eo}\oplus{\cal H}_{oe}\oplus{\cal H}_{oo}\ ,
   \]
where the subscripts $\sigma=\{ee, eo, oe, oo\}$ denote the even $e$ or odd $o$ parity of ${\Pi}_1$ and ${\Pi}_2$, respectively.

We use basis states labeled by $\vert \nu_{12}, \nu_{13}, \nu_{23} \rangle \otimes \vert n_1, n_2, n_3 \rangle$ with $n_1+n_2+n_3=N_{a}$ and $\nu_{jk} =0, 1,\cdots, \infty$. Because in our model a system of $3$-level atoms interacting with a $2$-mode field in a cavity generates an infinite dimensional Hilbert space, we need a truncation criterion to study the eigensystem of the Hamiltonian. For this purpose, we request convergence of the fidelity between base states $\vert\,\psi(k_{1max}, k_{2max})\rangle$ and $\vert\,\psi(k_{1max}+2, k_{2max}+2)\rangle$, where $(k_{1max}, k_{2max})$ are the maximum eigenvalues taken by the operators $\mathbf{K}_{1}$ and $\mathbf{K}_{2}$ in the current approximation. For the purposes of this work, we choose this convergence to be good when we reach an error of the order ${\rm e}_{\rm rr}=10^{-10}$, i.e.,
   \begin{equation}
	  1 - \mathcal{F}(k_1,k_2)\leq 10^{-10}\ .
	  \label{fidelityerror}
   \end{equation}
where $\mathcal{F}(k_1,k_2) = \vert\,\langle\psi(k_{1}, k_{2}) \vert\psi(k_{1}+2, k_{2}+2)\rangle\,\vert^2$ is the fidelity between the states. 
This fidelity constraint may be set according to the problem to be approached. We choose the approximation given in Eq.~(\ref{fidelityerror}) because it allows an approximation of the expectation value of the energy of the ground state good up to $10^{-8}$~\cite{cordero19,cordero19b}, even for large values of the coupling constants. So, in each Hilbert subspace ${\cal H}_{\sigma}$, the truncated basis ${\cal B}_\sigma$ is formed by the set of states $\vert \nu_{12}, \nu_{13}, \nu_{23} \rangle \otimes \vert n_1, n_2, n_3 \rangle$ with all eigenvalues $k_1\leq k_{1max}$ and $k_2\leq k_{2max}$ of the operators $\op{K}_1$ and $\op{K}_2$ respectively, and which preserve the parity $\sigma$. 
The basis ${\cal B}_\sigma$ obtained in this form will be called the {\it exact basis}.

\section{Fidelity as Signature of Quantum Phase Transitions}
\label{fid-qpt}

   \begin{figure*}
	\begin{center}
		\includegraphics[width=0.31\linewidth]{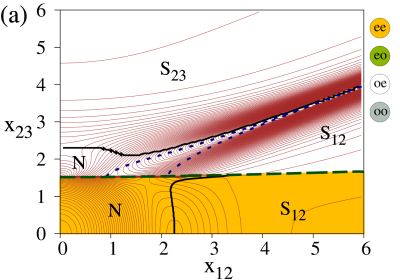}
		\includegraphics[width=0.31\linewidth]{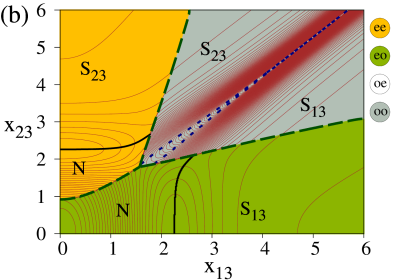}
		\includegraphics[width=0.31\linewidth]{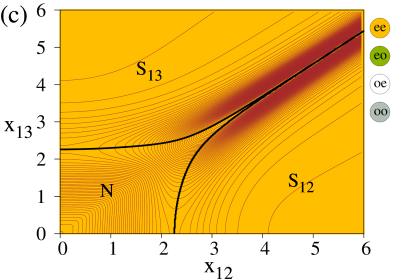}
	\end{center}
\caption{(colour online) Quantum separatrix for a single $3$-level atom interacting dipolarly with two modes of electromagnetic field, a generalised quantum Rabi model. (a) $\Xi$-configuration with parameters $\Omega_{12}=1/4$, $\Omega_{23}=3/4$ and $\omega_2=~1/4$. (b) $\Lambda$-configuration with parameters $\Omega_{13}=~1$, $\Omega_{23}=~9/10$, and $\omega_2=~1/10$. (c) $V$-configuration with parameters $\Omega_{12}=4/5$, $\Omega_{13}=1$ and $\omega_2=~4/5$. In all case we have fixed $\omega_1=~0$ and $\omega_3=~1$.}
\label{quant_sep}
\end{figure*}

Traditionally, to determine the quantum phase transitions in the limit $N_a \to \infty$, one uses a variational test function and calculates the associated ground state energy surface. This is an analytic function depending on parameters and variables which allows the calculation of its minimum critical points. Those associated to the so-called degenerate critical points determine the locus of points in parameter space where the ground state suffers a sudden change in its properties. Additionally, one can use the Ehrenfest classification, which determines the order of the quantum phase transitions according to the order of the derivative of the energy surface where the analyticity is lost~\cite{castanos12b}.

Here we are considering {\it finite} quantum systems, that is, systems with a finite number of particles. The ground state energy is calculated by means of the diagonalisation of the corresponding Hamiltonian matrix, and in order to determine the regions where a sudden change in the characteristics or properties of the ground state takes place we use quantum information concepts, such as the fidelity and the susceptibility of the fidelity. We shall here refer to such a  change in the behaviour and constitution of the ground state as a quantum phase transition; in the current literature it has also been called a quantum crossover.

The loci where the fidelity between neighbouring states $|\Psi_g({\xi_1})\rangle$, $|\Psi_g({\xi_2})\rangle$ along parametric lines $\xi(t)$ in parameter space
\begin{equation}\label{eq.fide}
{\cal F}(\rho_{\xi(t)},\rho_{\xi(t+\delta)}) = |\langle \Psi_g({\xi(t)})|\Psi_g({\xi(t+\delta)})\rangle|^2\,
\end{equation}
presents a minimum, determine points of the separatrix. For a large number of particles, one may follow trajectories which are parallel to the axes, as all other trajectories will yield the same separatrix (cf. Fig.~\ref{phasediagram}).  However, for a finite number of particles, one must consider trajectories in all directions of the plane $(x_{ij},\,x_{jk})$. Thus, in order to obtain a fine description of the phase diagram we consider the surface of minimum fidelity, calculated by considering neighbouring points in directions parallel to the axes ($x_{jk}=0$), along identity lines ($x_{ij}=x_{jk}$), and along their orthogonal directions ($x_{ij}=-x_{jk}$), thereby finding the local minima (see~\ref{ap.supfide}).

For the three-level systems interacting with two radiation modes in a cavity, in the case of $N_a=1$ particle (i.e., a generalised quantum Rabi model [cf.~\cite{braak16} and references therein]), by means of the fidelity and the susceptibility of the fidelity we have found three types of loci of points where the ground states changes abruptly (cf. figure~\ref{quant_sep}): The dashed lines indicate discontinuous transitions where the fidelity Eq.~(\ref{eq.fide})  of neighbouring states falls to zero, indicating that the neighbouring states in question are completely dissimilar. The separatrix in this case borders along orthogonal Hilbert subspaces of different parity.  On the other hand (full and dotted lines in the figure), there are situations with $F(\xi)\neq 0$ and it either remains different from zero as $N_a$ increases, or reaches zero in the large $N_a$ limit; these both are continuous transitions which we propose to call {\it stable} and {\it unstable} continuous transitions; they can be distinguished by means of the calculation of the Bures distance~\cite{bures69, helstrom67}, which measures the difference of two probability densities of the quantum system: if $\rho_A$ and $\rho_B$ are the density matrices of states $\vert\Psi_A\rangle$ and $\vert\Psi_B\rangle$, the Bures distance between the states is given by
\begin{equation}\label{eq.bures}
	D_B{}^2(\rho_A,\rho_B) = 2\left( 1 - \sqrt{{\cal F}(\rho_A,\rho_B)} \right),
\end{equation}
where the fidelity may be calculated as
\begin{equation}
	{\cal F}(\rho_A,\rho_B) = \left[ {\rm Tr}\,\sqrt{\sqrt{\rho_A} \rho_B \sqrt{\rho_A}} \right]^2 ,
\end{equation}
which reduces to ${\cal F}(\rho_A,\rho_B)=\vert\langle\Psi_A\vert\Psi_B\rangle\vert^2$ for pure states.

For the stable continuous transition the value of the Bures distance is smaller than for the unstable continuous transition~\cite{cordero20}.

The first order quantum phase transitions, according to the Ehrenfest classification, can be always determined by means of the Hellmann-Feynman theorem~\cite{hellmann37, feynman39}. These type of transitions are always associated to zero fidelity values, that is, discontinuous transitions. On the other hand, the associated changes in the properties of the ground state of the system will be clearly shown in the calculation of the Wigner quasi-probability distribution function of the two-modes of the electromagnetic field in the cavity. These results distinguish the three types of quantum phase transitions appearing for finite quantum systems, viz., the discontinuous, continuous-stable, and continuous-unstable quantum phase transitions.

For few particles (in this work we consider only $1$ particle, the generalised Rabi Model), we use the exact basis introduced in the previous section to calculate the ground state of the system. In order to obtain each minimum energy surface in phase space we use the fidelity criterion given in Eq.~(\ref{fidelityerror}), for each pair of symmetry values $k_{1{\rm max}}$ and $k_{2{\rm max}}$ and comparing these with values up to $(k_{1{\rm max}}+2)$ and $(k_{2{\rm max}}+2)$ (for the exact procedure cf.~\cite{cordero19,cordero19b}). 

The separatrix for the $\Xi-$, $\Lambda-$, and $V-$configurations are those shown in Fig.~(\ref{quant_sep}), where the parity of the eigenvalues of the operators $\op{K}_1$ and $\mathbf{K}_2$ Eq.~(\ref{eq.Kj3l}), preserved for each energy surface, is indicated by the coloured region and legend. The shape of the surface of maximum Bures distance for each atomic configuration is found in~\ref{ap.supfide}, where the discontinuities and local maxima correspond perfectly with the separatrices in figure~\ref{quant_sep}.

For the $\Xi$ configuration, figure \ref{quant_sep}(a), we see that the parity of the ground state of the system can have only two possibilities, $ee$ and $oe$, depending on the values of the coupling parameters $x_{ij}$. Outside the Normal region, the bi-modal character of light is present because mode $\Omega_{23}$ is dominant above the dashed line while below it the mode $\Omega_{12}$ is preponderant. The fine classification yields stable-continuous transitions (solid line) in the $ee$ and $oe$-regions and unstable-continuous transitions (dotted line) in the $oe$-region.

For the $\Lambda$ configuration, figure \ref{quant_sep}(b), the energy surface is formed by three parity regions $ee$, $eo$ and $oo$. Stable-continuous transitions (solid lines) occur in the $ee$- and $eo$-regions, while unstable-continuous transitions (dotted line) are had in the $oo$-region.

For the $V$ configuration, figure \ref{quant_sep}(c), a comparison with its thermodynamic counterpart in Fig.~\ref{phasediagram} shows clearly the correspondence with the domains where each electromagnetic mode is dominant. The ground energy surface has parity $ee$, the separatrix (solid line) presents stable-continuous transitions except when the curves coalesce, at which points unstable-continuous transitions occur.

The finer classification of the continuous transitions is more evident through the study of the quasi-probabilities, since this classification is based on whether the bulk of the ground state remains in a sub-basis of the total basis or not. In this work we focus our study on the properties of the Wigner function in the different regions. As the $\Lambda$ configuration appears to have a richer structure, which we will consider it in what follows, and in section~\ref{quantum-transitions} we will discuss its separatrix at length. The Supplementary Material contains the results for all atomic configurations.

\section{Calculation of the Wigner Function }
\label{Wigner-calc}

In order to study the quantum phase transitions, we make use of the Wigner function of the electromagnetic modes. In this section we calculate the Wigner function following the procedure outlined in~\cite{castanos18}. We denote the Fock basis states for the $\Lambda$-configuration by
   \begin{equation}
      \vert \nu_{13},\,\nu_{23},\,n_{1},\,n_{2},\,n_{3}\rangle\ ,
   \end{equation}
with the first two labels denoting the electromagnetic quanta oscillations number, and the next three the population of the atomic levels, which satisfy $n_{1}+n_{2}+n_{3}=N_{a}$. For this configuration one has the parity operators $e^{i\, \mathbf{K}_{1} \, \pi}$ and $e^{i\, \mathbf{K}_{2} \, \pi}$ with
 \begin{eqnarray}
       \mathbf{K}_{1}&=&\boldsymbol{\nu}_{13}+\boldsymbol{\nu}_{23}+\mathbf{A}_{33}\ ,\\
       \mathbf{K}_{2}&=&\boldsymbol{\nu}_{23}+\mathbf{A}_{11}+\mathbf{A}_{33}\ ,
   \end{eqnarray}
whose eigenvalues we denote by $k_{1}$ and $k_{2}$, respectively. We may use these to replace the electromagnetic quanta oscillations numbers,
   \begin{equation}
      \nu_{13}=k_{1}-k_{2}+n_{1}\ ,\qquad \nu_{23}=k_{2}-n_{1}-n_{3}\ ,
   \end{equation}
and thus denote the ground state of the system as

\begin{eqnarray*}
    && \vert\psi_{\rm gs}\rangle= \sum_{k_{1},k_{2}} \sum^{N_{a}}_{n_{1},n_{3}} C_{k_{1},k_{2},n_1,n_3} \\
    &&\times\vert k_{1}-k_{2}+n_{1}, k_{2}-n_{1}-n_{3},\,  n_{1}, \, N_a-n_1-n_{3},\,  n_{3} \rangle  \end{eqnarray*}
Notice that for the Tavis-Cummings model we do not have the sum over indices $k_{1}$, $k_{2}$, as these are associated to constant of the motion. For the Dicke model, although $k_{1}$, $k_{2}$ are not fixed, their parity is invariant.

The density matrix of the ground state of the system can be calculated from the expression above, and from it the reduced density matrices for the modes $\nu_{13}$ and $\nu_{23}$ are obtained:

   \begin{eqnarray*}
      \mathbf{\varrho}_{13}&=&\sum_{k_{1},k_{1}^{\prime},k_{2}}\sum_{n_{1},n_{3}}
      C_{k_{1},k_{2},n_1,n_3}\,C_{k_{1},k_{2}^{\prime},n_1,n_3}^{\ast}\\
      &&\times\,\vert k_{1}-k_{2}+n_{1}\rangle\langle k_{1}^{\prime}-k_{2}+n_{1}\vert\ ,\\
      && \\   
      \mathbf{\varrho}_{23}&=&\sum_{k_{1},k_{2},k_{2}^{\prime}}\sum_{n_{1},n_{3}}
      C_{k_{1},k_{2},n_1,n_3}\,C_{k_{1}^{\prime},k_{2},n_1,n_3}^{\ast}\\
      &&\times\,\vert k_{2}-n_{1}-n_{3}\rangle\langle k_{2}^{\prime}-n_{1}-n_{3}\vert\ .
\label{rhos}
   \end{eqnarray*}

In order to calculate the Wigner function of the system, one uses an expression for the Weyl symbol $W_{\vert n\rangle\langle m \vert}(q,p)$ of the operator $\rho_{nm}=\vert n\rangle\langle m\vert$. Writing the Glauber coherent state in the position representation, one arrives at the normalised expression for the Wigner function
\begin{eqnarray}
W_{\alpha,\beta}(q,p) &=& \exp{\left\{ -\frac{\vert \alpha\vert^2}{2} - \frac{\vert \beta\vert^2}{2}  \right\}}  \nonumber \\
&\times& \, \exp{\left\{ -z \, z^*+ \sqrt{2} \, \alpha\,z^*  + \sqrt{2} \, \beta^* \,z -\alpha \, \beta^* \right\} }  \, ,\nonumber \\
&&
\label{wig2}
\end{eqnarray}
where we have defined the complex variable $z=q+ip$ and the function is normalised with respect to the volume element $ d\mu= dq\, dp$.

Considering the expansion of the coherent states $\vert \alpha\rangle$ and $\vert \beta\rangle$ with respect to Fock states we have
\begin{eqnarray*}
W_{\alpha,\beta}(q,p) &=& 
e^{-\frac{1}{2}(\vert \alpha\vert^2+ \vert \beta\vert^2)}
\sum_{n,m} \frac{\alpha^n \, \beta^{* m}}{\sqrt{n! \, m!} } \, W_{\vert n\rangle\langle m \vert}(q,p) \, .
\label{wig3}
\end{eqnarray*} 

Now, through the generating function for the associated Laguerre polynomials, and some algebra, we arrive at
\begin{eqnarray}
W_{\vert n\rangle\langle m \vert}(q,p)&=& \frac{(-1)^m }{\pi} \, 2^\frac{n-m}{2} \, \sqrt{\frac{m!}{n!}} \, (q- ip)^{n-m}\nonumber \\
&\phantom{=}&\times e^{-(q^2 + p^2)} \, L^{n-m}_m(2 (q^2+p^2))\ ,
\end{eqnarray}
for $n\geq m$. For $n < m$ we need to interchange $n \leftrightarrow m$, together with $q- ip \to q+ ip$.
In this manner we obtain the Wigner function for the reduced density matrices:

\begin{eqnarray}
	\fl
W_{13}(q,\,p)&=&\sum_{k_{1},k_{2},k_{1}^{\prime}}\sum_{n_{1},n_{3}}
C_{k_{1},k_{2},n_1,n_3}\,C_{k_{1}^{\prime},k_{2},n_1,n_3}^{\ast}
W_{\vert k_{1}-k_{2}+n_{1}\rangle\langle k_{1}^{\prime}-k_{2}+n_{1} \vert}(q,p)\ ,\\
\fl
W_{23}(q,\,p)&=&\sum_{k_{1},k_{2},k_{2}^{\prime}}\sum_{n_{1},n_{3}}
C_{k_{1},k_{2},n_1,n_3}\,C_{k_{1},k_{2}^{\prime},n_1,n_3}^{\ast}
W_{\vert k_{2}-n_{1}-n_{3}\rangle\langle k_{2}^{\prime}-n_{1}-n_{3} \vert}(q,p) \ .
\end{eqnarray}

\section{Wigner function and quantum phase transitions}
\label{quantum-transitions}

\begin{figure}
	\begin{center}
	\includegraphics[width=0.6\linewidth]{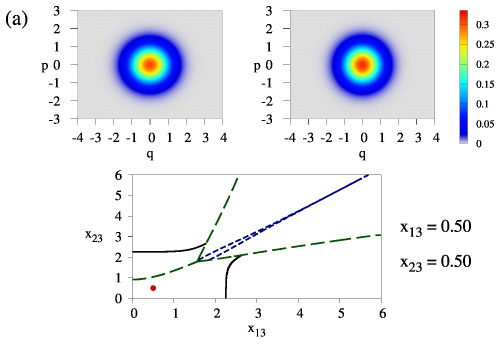}\\
		{\includegraphics[width=0.6\linewidth]{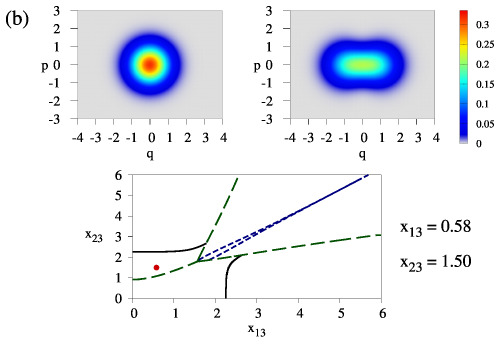}}\\
		{\includegraphics[width=0.6\linewidth]{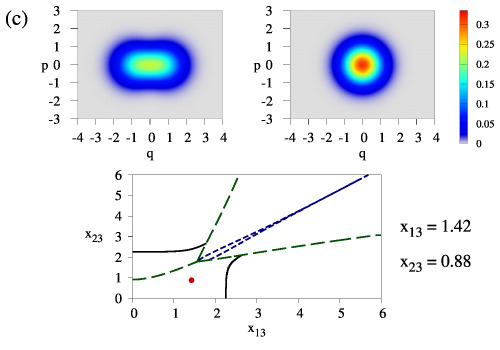}}
	\end{center}
	 \caption{(Colour online) behaviour of the Wigner function in the normal region for (a) a point with small values of both $x_{13}$ and $x_{23}$, (b) a point in the region $x_{13}<x_{23}$ over the separatrix, and (c) a point the region $x_{13}>x_{23}$ below the separatrix. The values of $x_{13}$ and $x_{23}$ at the point in question (red in the figure) are indicated.}
\label{wignerLNa1a}
\end{figure}

\begin{figure}
	\begin{center}
		{\includegraphics[width=0.6\linewidth]{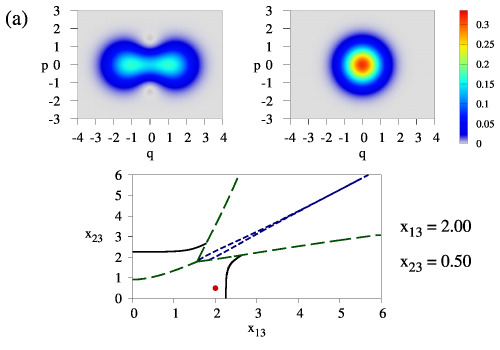}}\\
		{\includegraphics[width=0.6\linewidth]{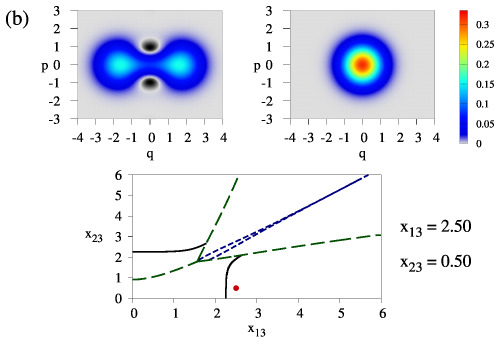}}
	\end{center}
	 \caption{(Colour online) behaviour of the Wigner function as the system goes through a stable-continuous transition.}
\label{wignerLNa1b}
\end{figure}

\begin{figure*}
	\begin{center}
		{\includegraphics[width=0.45\linewidth]{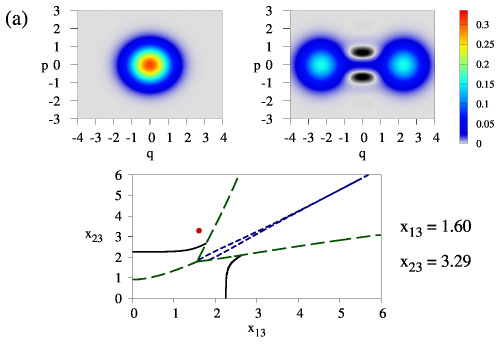}} 
		{\includegraphics[width=0.45\linewidth]{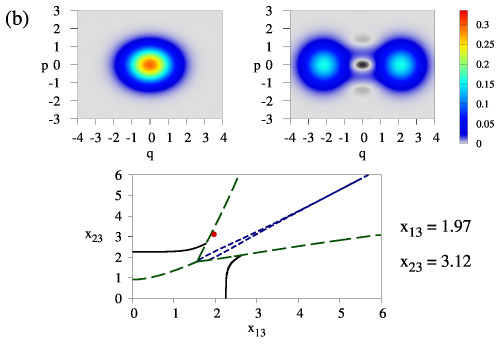}}\\ 
		{\includegraphics[width=0.45\linewidth]{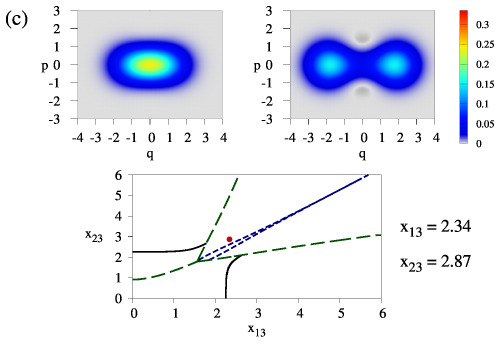}} 
		{\includegraphics[width=0.45\linewidth]{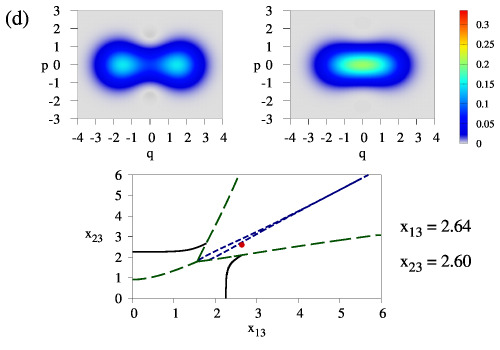}}\\
		{\includegraphics[width=0.45\linewidth]{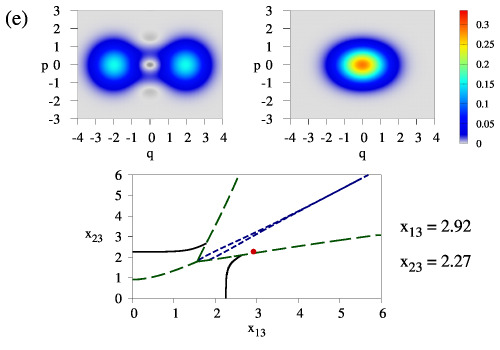}} 
		{\includegraphics[width=0.45\linewidth]{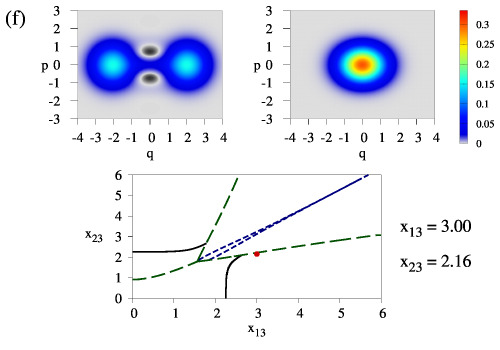}} 
	\end{center}
	 \caption{(Colour online) behaviour of the Wigner function close to the normal region as the system goes through different phase transitions: (a)-(b) a discontinuous transition $ee\rightleftharpoons oo$, (c)-(d) an unstable-continuous transition, and (e)-(f) a discontinuous transition $oo\rightleftharpoons eo$. The corresponding separatrices according to the type of transition, and the points of calculation (red in the figure), are indicated.}
\label{wignerLNa1c}
\end{figure*}

\begin{figure*}
	\begin{center}
		{\includegraphics[width=0.45\linewidth]{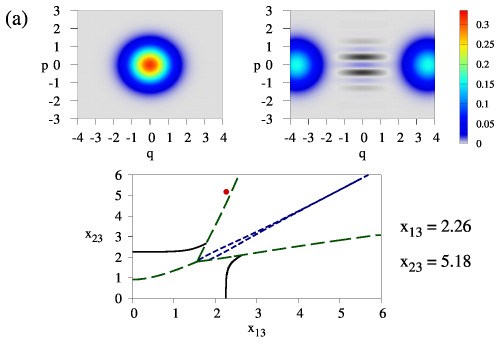}}
		{\includegraphics[width=0.45\linewidth]{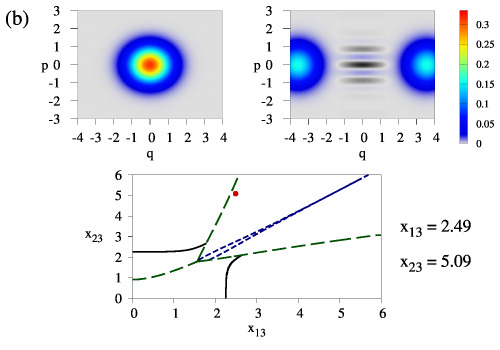}}\\
		{\includegraphics[width=0.45\linewidth]{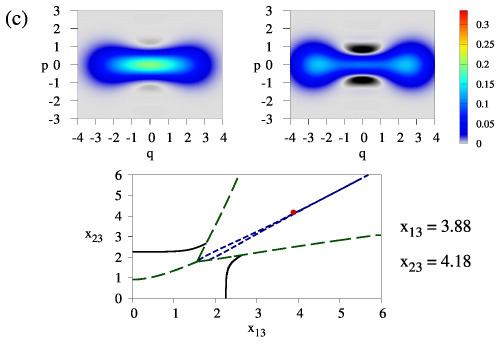}}
		{\includegraphics[width=0.45\linewidth]{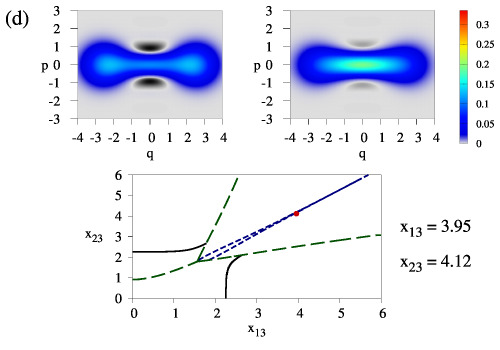}}\\
		{\includegraphics[width=0.45\linewidth]{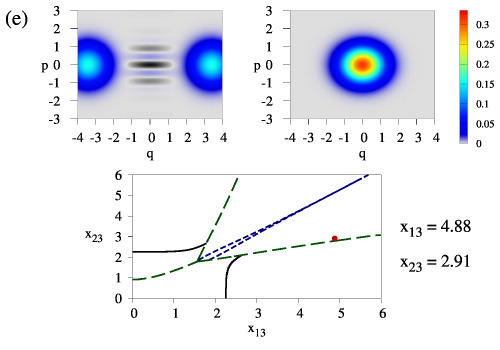}}
		{\includegraphics[width=0.45\linewidth]{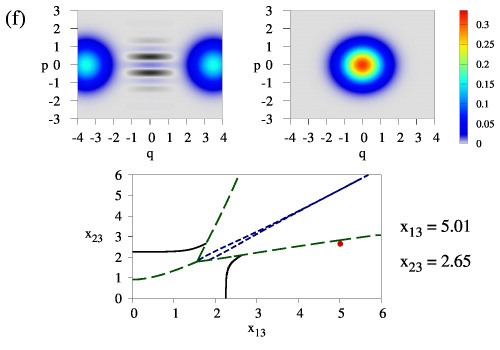}}
	\end{center}
	\caption{(Colour online) behaviour of the Wigner function for large values of the dipolar strengths, as the system goes through different phase transitions: (a)-(b) a discontinuous transition $ee\rightleftharpoons oo$, (c)-(d) an unstable-continuous transition, and (e)-(f) a discontinuous transition $oo\rightleftharpoons eo$. The corresponding separatrices according to the type of transition, and the points of calculation (red in the figure), are indicated.}
\label{wignerLNa1e}
\end{figure*}

In the following figures we illustrate the quantum phase transitions for $N_{a}=1$ atom in the $\Lambda$-configuration, with the parameters given in Fig.~\ref{quant_sep}. We plot the Wigner function as a function of the quadratures $q$ and $p$ at various points at either side of a separatrix (continuous, dashed and dotted lines in the figures), in order to show how it behaves as the system undergoes a phase transition. Each plot displays $W_{13}$ at the upper right and $W_{23}$ at the upper left, the phase diagram is shown below them, and in it a solid (red) dot marks the point at which the Wigner function is calculated. We also provide, as supplementary material, movies which show, for the three atomic configurations $\Xi$, $\Lambda$, and $V$, the Wigner function along a trajectory which goes through different regions of the phase diagram crossing the different separatrices.

We first note that the phase diagram for $N_{a}=1$ reveals a rich structure not present in the thermodynamic limit of the system. In the thermodynamic limit, there are only three regions present: the normal regime, where the behaviour is not dominated by any of the electromagnetic modes, and two collective regions where one of the electromagnetic modes is prevalent [cf. figure~\ref{phasediagram}(b)]. On the other hand, the phase diagram for $N_{a}=1$ divides the parameter space into three regions with fixed parity, each of which in turn splits into at least two regions [cf. figure~\ref{quant_sep}(b)].

In Fig.~\ref{wignerLNa1a}(a), the Wigner functions $W_{13}$ and $W_{23}$ are shown for small values of the parameters $x_{13}$ and $x_{23}$. Both functions are positive and have circular symmetry, i.e., the bulk of the ground state is dominated by the vacuum state of the field. Notice that the normal region is divided into two regions by a separatrix (dashed line) where a change of parity occurs (discontinuous transition). Thus, it is of interest to consider points above and below this separatrix, at points where the small contribution of the radiation field is not negligible. This is shown in Figs.~\ref{wignerLNa1a}(b) and \ref{wignerLNa1a}(c). As we move from one side of the separatrix to the other, the Wigner function of one mode passes from a circularly symmetric shape to an elongated one along the $q-$axis, while that of the other mode does the opposite. The fidelity criterion gives us a discontinuous phase transition, since a change of parity occurs. From the Wigner function point of view, the bulk of the ground state changes from a subset of the basis with a major contribution from one kind of photons, to a subset with a major contribution of the other one [compare $W_{13}$ and $W_{23}$ in figures~\ref{wignerLNa1a}(b) and \ref{wignerLNa1a}(c)]. This behaviour of the Wigner function, which detects this transition in the normal region, reflects the fact that the bulk of the ground state is inside a Hilbert subspace corresponding to a $2$-level subsystem, similar to the variational solution.

Fig.~\ref{wignerLNa1b} depicts the behaviour of the Wigner function as the system goes through a stable-continuous transition (solid line), for small values of  $x_{23}$. In both subregions $W_{13}$ elongates along the $q-$axis while $W_{23}$ remains without change. (Strictly speaking, the only contribution of field states of mode $\Omega_{23}$ is the vacuum). We can see that regions where the Wigner function $W_{13}$ is negative appear as we move away from the normal region and cross the separatrix, a sign of quantumness of the ground state. This is because the number of photons in mode $\nu_{13}$ grows from zero, and we now have a superposition of states with different values of $\nu_{13}$. We will see below that when $x_{13}$ increases, the region between the two main bulks which constitute $W_{13}$ grows as well, and the black (negative valued) regions also grow, reflecting the fact that we have more different values of $\nu_{13}$ in superposition. This is a sign of entanglement in the system. Similar results are obtained when the other separatrix (solid line) is crossed (with small values of $x_{13}$ and growing $x_{23}$); in this case the mode $\Omega_{23}$ dominates.

Figures~\ref{wignerLNa1c} and \ref{wignerLNa1e} show, respectively, what occurs when one moves across separatrices in the collective region, for small and large values of the dipolar strengths.  In this collective region one finds two discontinuous transitions (dashed lines) due to the change of parity $ee\rightleftharpoons oo$ and $oo\rightleftharpoons eo$ in the ground state, and an unstable-continuous transition (dotted line) which occurs when the state has $oo$ parity [please refer back to Fig.~\ref{quant_sep}(b)].

Discontinuous transition  $ee\rightleftharpoons oo$: the Wigner function for the mode $\Omega_{13}$ in the collective region $ee$ (above the solid line and to the left of the dashed line) is qualitatively equal to that of the vacuum state, having circular symmetry with positive values, i.e., the contribution of photons $\nu_{13}$ is negligible in this region, while the photon contribution $\nu_{23}$ is significant, reflected by the fact that $W_{23}$ is very elongated, presenting a bimodal distribution and having negative values (black regions). A similar behaviour of the Wigner function occurs in the region $oo$ once the transition takes place. One may observe that  $W_{13}$ does not detect the discontinuous transition $ee\rightleftharpoons oo$, while $W_{23}$ shows a strong change of phase, taking now negative values at the origin [cf. Figs.~\ref{wignerLNa1c}(a)-\ref{wignerLNa1c}(b) and the corresponding Figs.~\ref{wignerLNa1e}(a)-\ref{wignerLNa1e}(b)]. This change of phase in the Wigner function may be see as a change from a male to a female Schr\"odinger cat of the field~\cite{castanos95}.

Unstable-continuous transition [Figs.~\ref{wignerLNa1c}(c)-\ref{wignerLNa1c}(d) and Figs.~\ref{wignerLNa1e}(c)-\ref{wignerLNa1e}(d)]: close to the separatrix in dotted lines both photon contributions are significant. Both Wigner functions present elongated (bimodal) distributions. Above the separatrix the contribution of photons $\nu_{23}$ dominates ($W_{23}$ has major regions with negative values), while $\nu_{13}$ dominates in the region below the separatrix. The continuous transition is smoother where the separatrix is clearly bifurcated [cf. Figs.~\ref{wignerLNa1c}(c)-\ref{wignerLNa1c}(d)] and more abrupt when it is not [cf. Figs.~\ref{wignerLNa1e}(c)-\ref{wignerLNa1e}(d)]. In both cases, however, when an unstable-continuous transition occurs the field mode contributions to the ground state change their roles.

Discontinuous transition $oo\rightleftharpoons eo$  [Figs.~\ref{wignerLNa1c}(e)-\ref{wignerLNa1c}(f) and Figs.~\ref{wignerLNa1e}(e)-\ref{wignerLNa1e}(f)]: this is the dual of the previous discontinuous transition. The contribution of the mode $\Omega_{23}$ is now negligible, while the state of photons of type $\nu_{13}$ is the one suffering a change of phase, in a similar fashion to the case of the discontinuous transition $ee\rightleftharpoons oo$ discussed above.

From the results above we see that the Wigner function characterises completely the phase diagram. In the normal region, the Wigner function describes a classical behaviour of the field ($W$ takes positive values) and at least one photon mode remains in the vacuum (cf. Fig.~\ref{wignerLNa1a}). The collective region is characterised by a Wigner function in which the quantumness of the photon modes is clearly shown, or both photon contributions are significant. This latter case occurs close to the unstable-continuous transition (cf. Figs.~\ref{wignerLNa1c} and \ref{wignerLNa1e}). In addition, as in the variational solution shown in Fig.~\ref{phasediagram}(b), the collective region divides itself into two regions, in each of which a single radiation mode dominates.

A video which shows the contour plots of the Wigner functions for the two electromagnetic modes is available online as Supplementary Material~\cite{videos}. The trajectory in parameter space was chosen to illustrate all the phase transitions, and the behaviour of the system in the various regions.

\section{Correlations between the Wigner Function and Entanglement}
\label{linearentropy}

The bimodality and negativity of the Wigner function reflect which field mode dominates in the superradiant region, and not the change of parity of the state. This is evident when we compare it with an entanglement measure, such as the linear entropy. When a change of parity is experienced, only the negativity pattern of the Wigner function is modified, as shown in figures~\ref{wignerLNa1c} and \ref{wignerLNa1e}, subfigures (a) and (b), and subfigures (e) and (f)).

Here we calculate the linear entropy $S_L = 1 - {\rm Tr}(\rho_{\rm reduced}^2)$ between the different subsystems, where $\rho_{\rm reduced}$ is the reduced density matrix for the subsystem considered. Thus, $S_{L_1} = 1 - {\rm Tr}(\rho_{\nu_1}^2)$ refers to the linear entropy measuring the correlation between field mode $1$ and the rest of the system (matter $+$ field mode $2$), $S_{L_2} = 1 - {\rm Tr}(\rho_{\nu_2}^2)$ refers to the linear entropy measuring the correlation between field mode $2$ and the rest of the system (matter $+$ field mode $1$), and lastly $S_{L} = 1 - {\rm Tr}(\rho_{\nu_1 \nu_2}^2)$ refers to the linear entropy measuring the correlation between matter and field modes $1$ and $2$. (The reduced density matrices shown here are calculated as described in section~\ref{Wigner-calc}.) We shall see that this complements the results obtained through the Wigner function, and correlates with them very well both qualitatively and in a quantitative way.

\begin{figure}
	\begin{center}
	\includegraphics[width=0.45\linewidth]{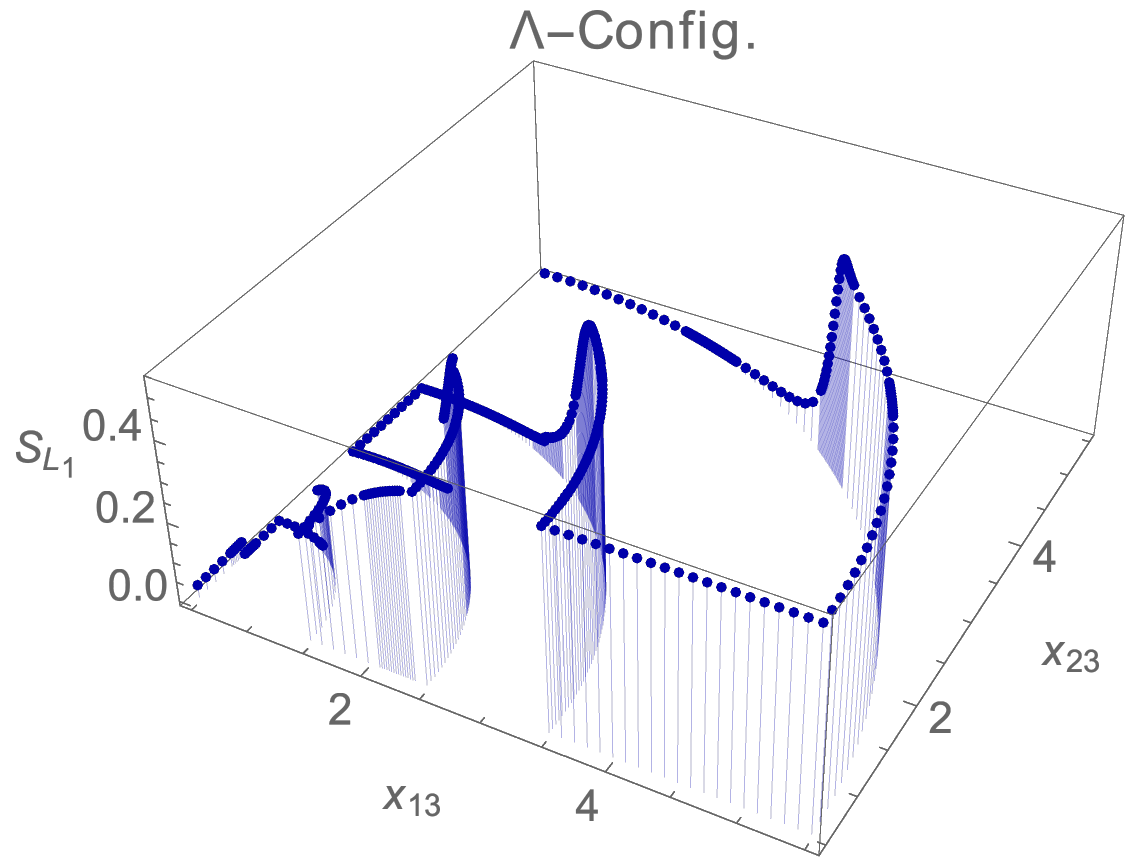}\quad
		{\includegraphics[width=0.45\linewidth]{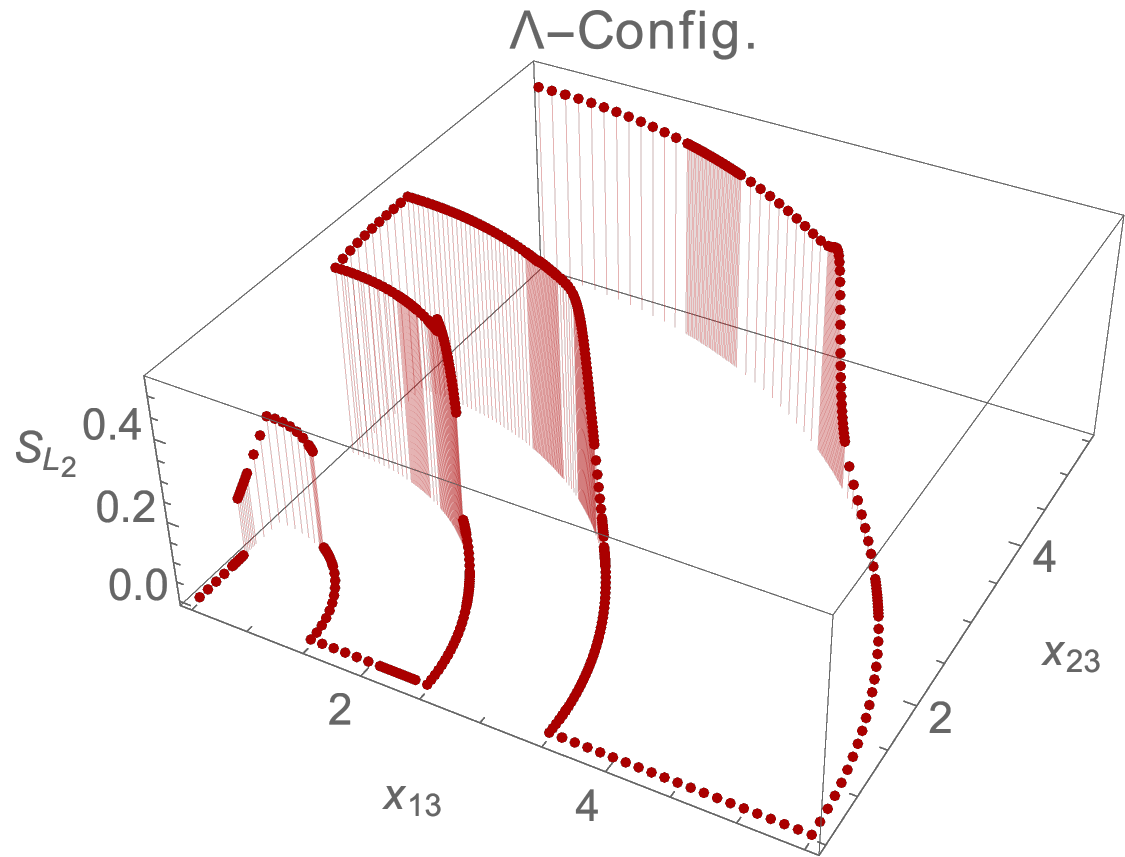}}\quad
		{\includegraphics[width=0.45\linewidth]{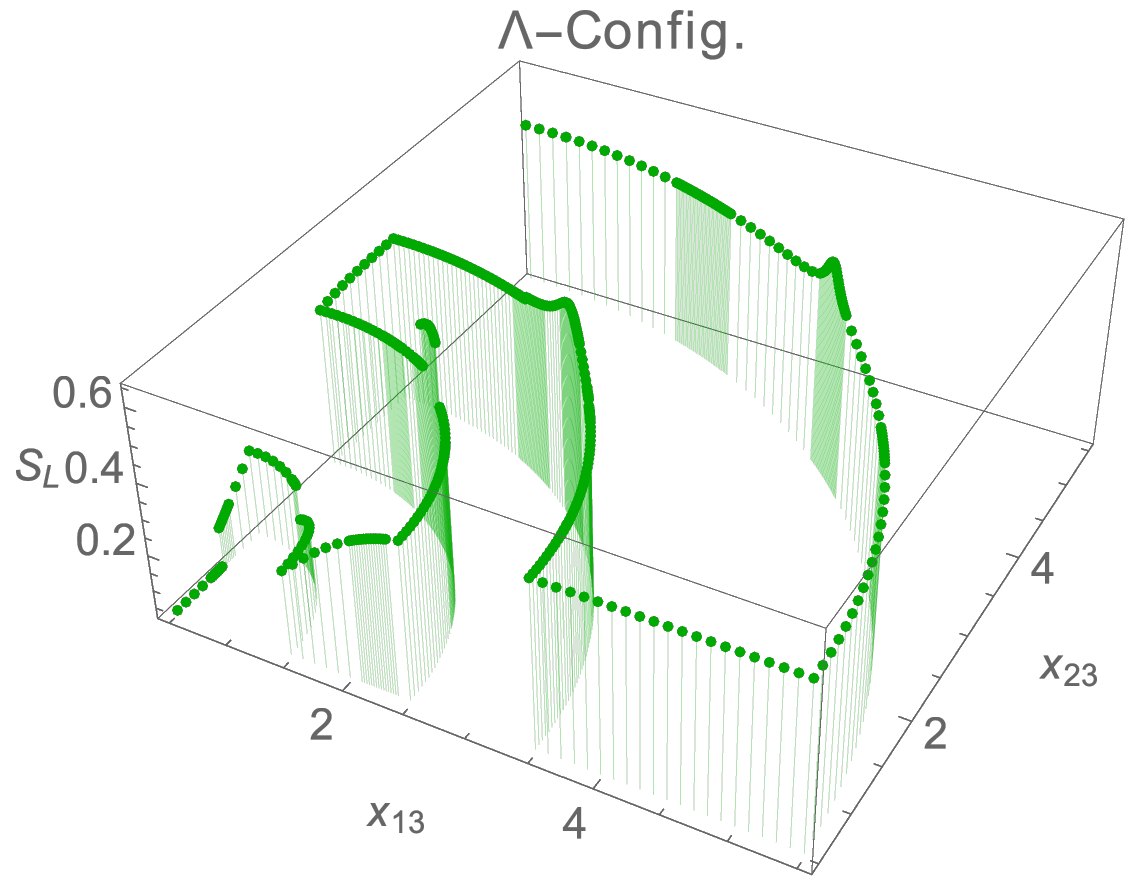}}
	\end{center}
	 \caption{(Colour online.) Linear entropy for the different subsystems, along a trajectory which crosses all detected transitions in the studied parameter space, for the generalised quantum Rabi model of $N_a=1$ atom in the $\Lambda$ configuration. The plot on the top left shows the correlation between field mode $1$ and the rest of the system, the plot on the top right the correlation between field mode $2$ and the rest of the system, and the plot on the bottom the correlation between matter and field modes $1$ and $2$. The density of calculated points was increased near the separatrices.}
\label{SLL}
\end{figure}

A comparison between figure~\ref{SLL}, for the atomic $\Lambda$ configuration, and the plots for the Wigner function in the previous section shows that, when the ground state is dominated by the vacuum state of the field (small values of the coupling parameters inside the Normal region), the correlation between one mode of the field, say $i$, and the rest of the system (matter $+$ field mode $j$ with $i \neq j$), is null $S_{L_i}=0$ and the Wigner function is unimodal. As we cross one of the separatrices inside the Normal region the linear entropy for the non-zero field mode $S_{L_i}$ starts to increase while the Wigner function elongates along one of the quadratures (depending on the non-null mode).

This field-mode $i$ vs. matter + field-mode $j$ entanglement reaches its maximum as soon as we cross into the superradiant region, the Wigner function showing negative values at a vicinity of the origin of quadrature $q$ and small non-zero values of quadrature $p$. This correlation remains maximum even when crossing from one parity region to another, as long as the field mode $i$ dominates the ground state, but falls rapidly to zero as soon as we enter the region where field mode $j$ dominates, even if a parity change is not had. The same is true for the Wigner function: its elongation and negativity reflect which field mode dominates in the superradiant region, and not  the change of parity of the state.

The linear entropy $S_L$, measuring the correlation between field (both modes) and matter, is always near its maximum value as soon as we enter the superradiant region, regardless of the parity of the ground state, and reaches its maximum values precisely at the continuous-unstable transition (dotted lines in Figure 3), even though the parity of the state is preserved as we cross it.

The same is true for the other two atomic configurations, $V$ and $\Xi$, shown in figures~\ref{SLV} and \ref{SLX} (compare with the Wigner functions presented in the Supplementary Material~\cite{videos}).

\begin{figure}
	\begin{center}
	\includegraphics[width=0.45\linewidth]{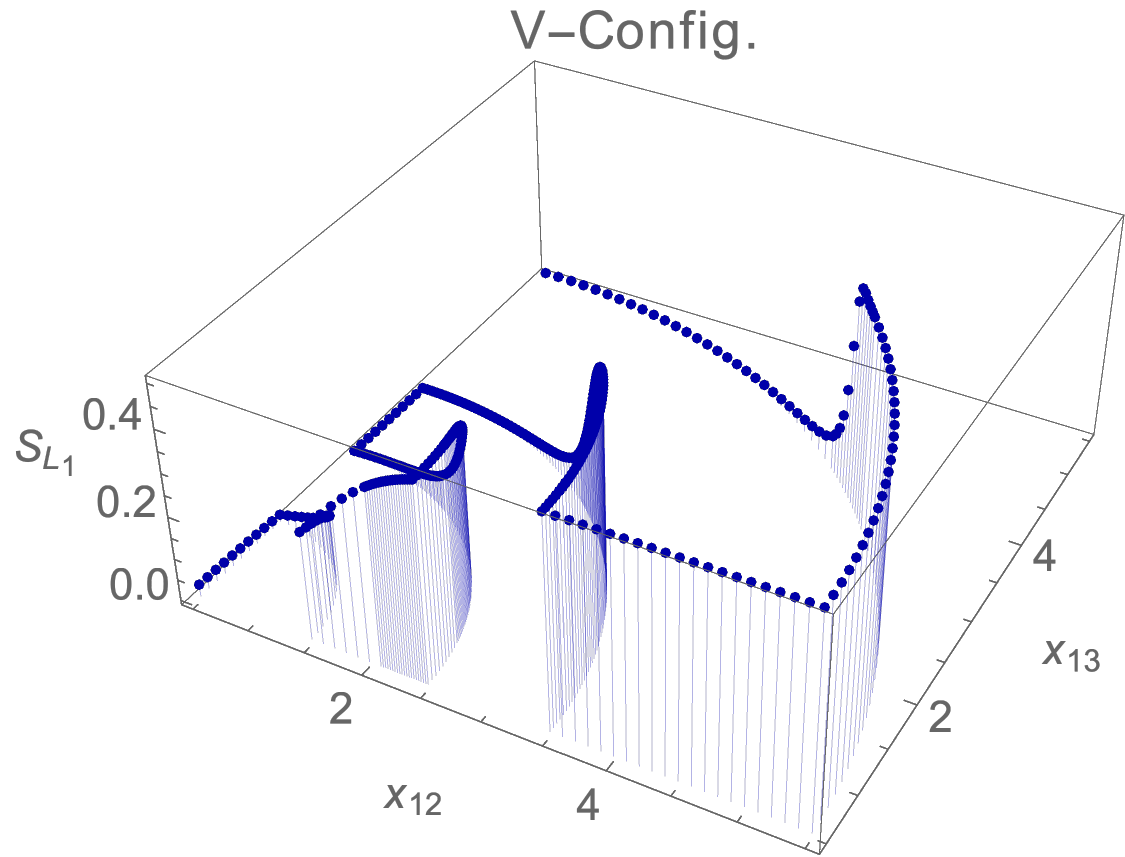}\quad
		{\includegraphics[width=0.45\linewidth]{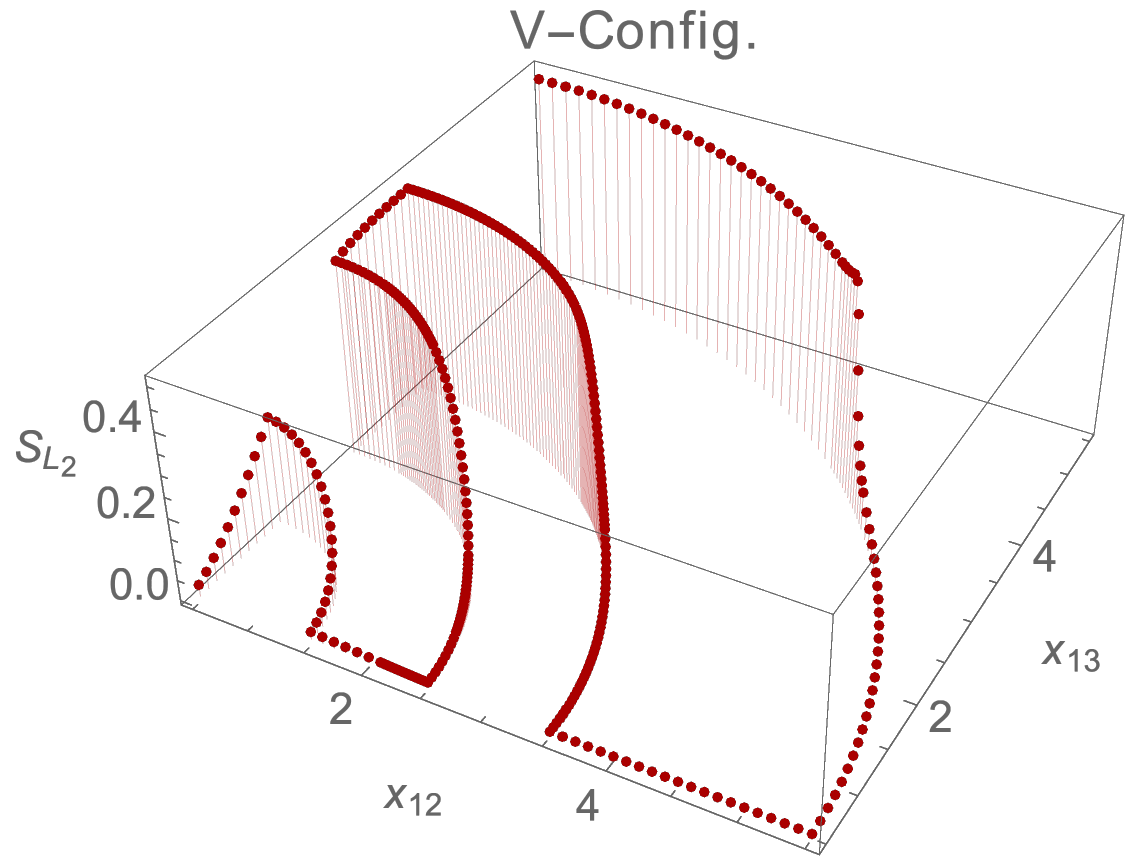}}\quad
		{\includegraphics[width=0.45\linewidth]{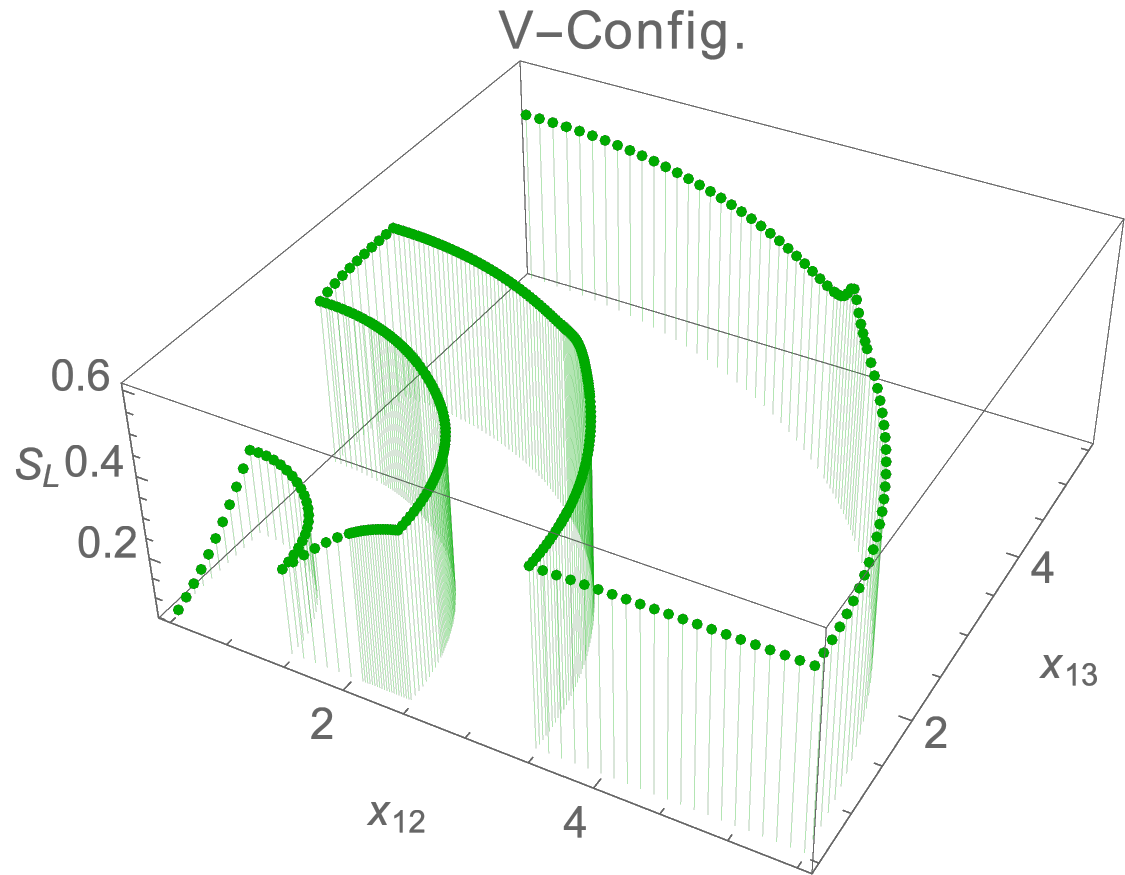}}
	\end{center}
	 \caption{(Colour online.) Linear entropy for the different subsystems, along a trajectory which crosses all detected transitions in the studied parameter space, for the generalised quantum Rabi model of $N_a=1$ atom in the $V$ configuration. The plot on the top left shows the correlation between field mode $1$ and the rest of the system, the plot on the top right the correlation between field mode $2$ and the rest of the system, and the plot on the bottom the correlation between matter and field modes $1$ and $2$. The density of calculated points was increased near the separatrices.}
\label{SLV}
\end{figure}

\begin{figure}
	\begin{center}
	\includegraphics[width=0.45\linewidth]{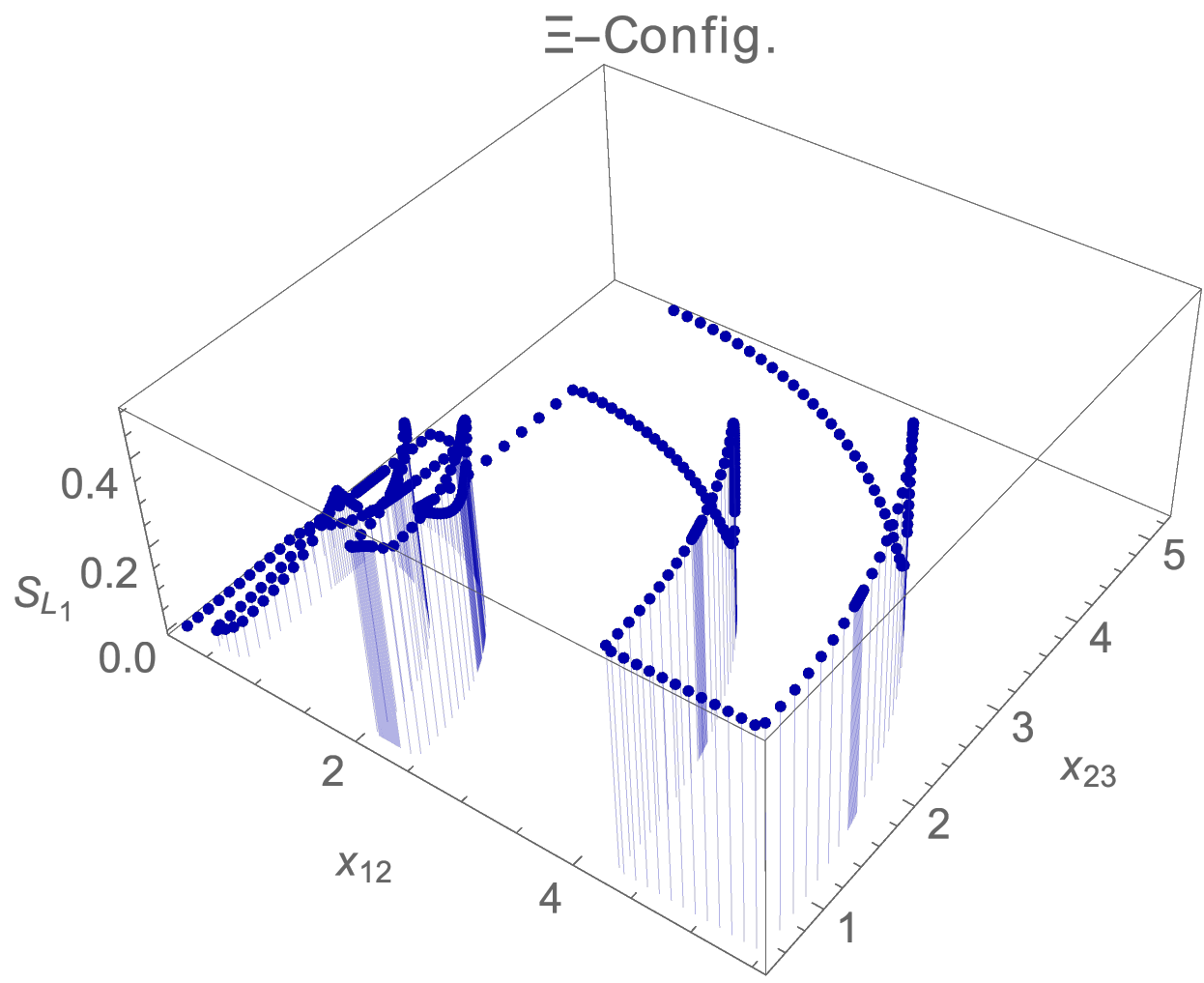}\quad
		{\includegraphics[width=0.45\linewidth]{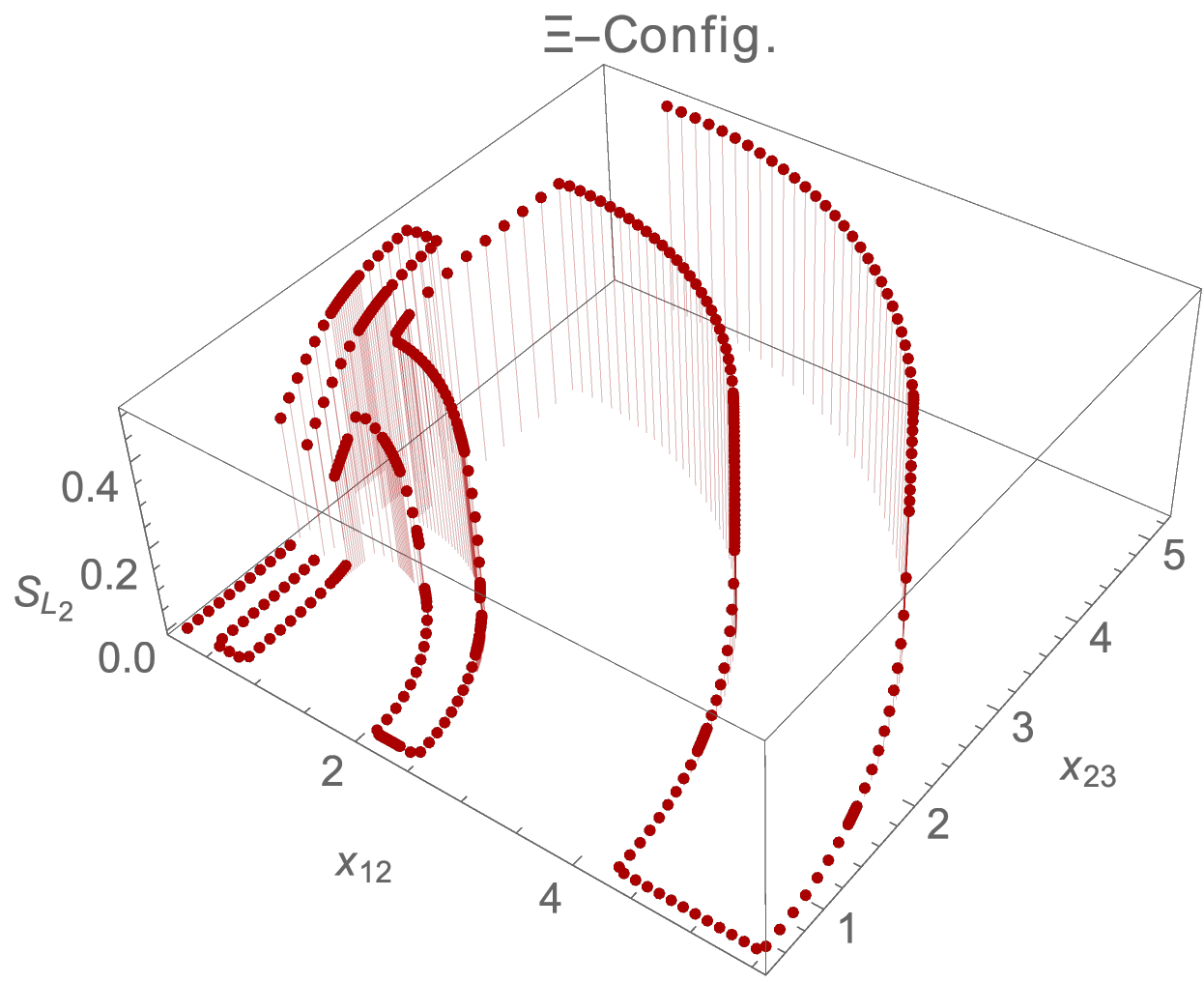}}\quad
		{\includegraphics[width=0.45\linewidth]{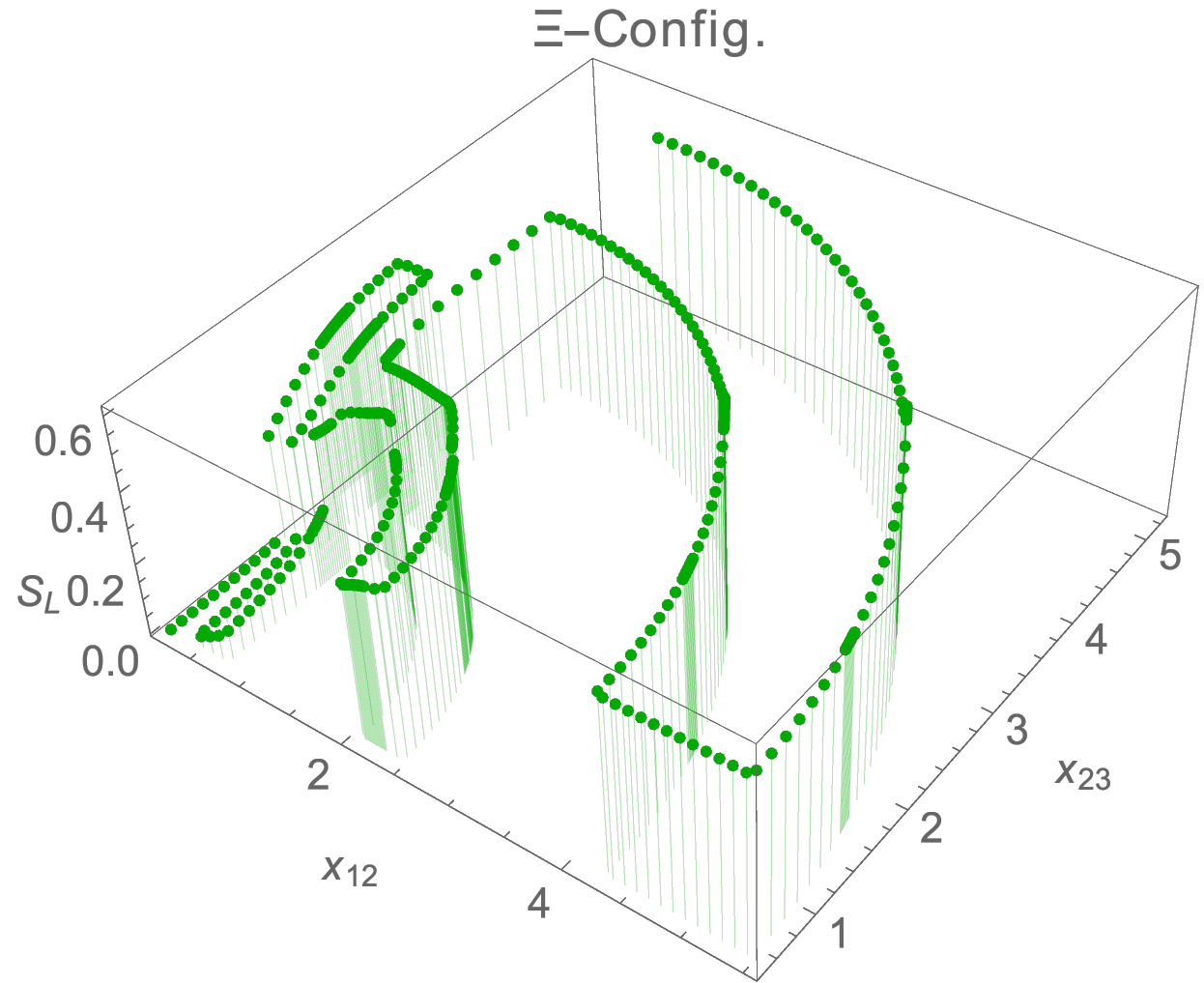}}
	\end{center}
	 \caption{(Colour online.) Linear entropy for the different subsystems, along a trajectory which crosses all detected transitions in the studied parameter space, for the generalised quantum Rabi model of $N_a=1$ atom in the $\Xi$ configuration. The plot on the top left shows the correlation between field mode $1$ and the rest of the system, the plot on the top right the correlation between field mode $2$ and the rest of the system, and the plot on the bottom the correlation between matter and field modes $1$ and $2$. The density of calculated points was increased near the separatrices.}
\label{SLX}
\end{figure}

So the entanglement between the substates responds to how the bulk of the ground state changes from a subset of the basis with a major contribution from one kind of photons, to a subset with a major contribution of the other one, and not to the state parity for large values of the coupling parameters. A change of parity produces a discontinuous phase transition, and may be detected through the fidelity criterion. In this sense, the study of the phase diagram through the Bures distance or fidelity criterion, and through the Wigner quasi-distribution function, are complementary.

\section{Conclusions}
\label{conclusions}

In this paper, we show the results of the characteristics of the ground state for a single three-level atom interacting dipolarly with a two-mode electromagnetic field. The symmetries of the system allow for the division the quantum state space into subspaces which have a well-defined parity with respect to these symmetries, which in turn reduces the dimension of the space considered to calculate the ground state. We have used a fidelity criterion to determine the quantum phase transitions for the three three-level configurations. The phase diagram of the $\Lambda$-configuration has the richest structure, and we discuss this case in detail, although the same reasoning can be followed for the other two configurations. To this end, we calculate the Wigner function for each of the electromagnetic modes $\Omega_{13}$ and $\Omega_{23}$, and show, in a series of plots, the behaviour of these Wigner functions in the various regions of the parameter space, which supplies further evidence of the quantum phase transitions revealed by the fidelity criterion~\cite{cordero20}. One important result is the determination of the regions where the ground state of a single atom in the $\Lambda$-configuration shows negative values in the Wigner function, because in these regions the system exhibits a non-classical behaviour. Through the use of the surface of maximum Bures distance, a finer classification is proposed for the continuous phase transitions. Lastly, the linear entropy for all the subsystems is calculated and, by comparing it with the behaviour of the Wigner function, we see that the entanglement between the substates responds to how the bulk of the ground state changes from a subset of the basis with a major contribution from one kind of photons, to a subset with a major contribution of the other one, and not to the state parity for large values of the coupling parameters.

\section*{Acknowledgments}
This work was partially supported by DGAPA-UNAM (under projects IN101619, IN112520, and IN100120).

\appendix
\section{Surface of Maximum Bures Distance}\label{ap.supfide}

As pointed out earlier, the phase diagram is determined by the set of points where a minimum of the fidelity (or a maximum of the Bures distance) between neighbouring states occurs. One may compute these quantities along a given trajectory in phase space because they depend on one parameter [cf. eq.(\ref{eq.fide})], e.g., one may consider variations of the fidelity along parametric curves, in parameter space, and determine its local minima; in particular, one may use linear trajectories. The result, however, will depend on the trajectory taken, and if we want to describe the behaviour of the Wigner function and relate its negativity to the entanglement of the subsystems, a finer description would be in order. 

In order to obtain a finer description of the phase diagram, and get rid of the dependance on the trajectory taken, we construct the {\em surface of minimum fidelity} ${\cal F}_{min}$ or, equivalently, since a minimum value of the fidelity yields also a maximum value of the Bures distance (\ref{eq.bures}), the {\em surface of maximum Bures distance} ${D_B}_{max}$, as follows:

Let $\delta A(\epsilon)$ be the neighbourhood of radius $\epsilon>0$ about a point $A$ in parameter space. Then, by considering the set of points of minimum fidelity or maximim Bures distance with respect to $A$, within that neighbourhood, and repeating the process for all points in parameter space, we obtain a surface of minimum fidelity, or the surface of maximum Bures distance 
\begin{equation}\label{eq.supfide}
{\cal F}_{min}(\rho_{A},\epsilon) = \min\left\{{\cal F}(\rho_{A},\rho_{B}) \big| B\in \delta A(\epsilon)\right\}\,.
\end{equation}
\begin{equation}\label{eq.supbures}
{D_B}_{max}(\rho_{A},\epsilon) = \max\left\{{D_B}(\rho_{A},\rho_{B}) \big| B\in \delta A(\epsilon)\right\}\,.
\end{equation}

Since constructions (\ref{eq.supfide}) and (\ref{eq.supbures}) do not depend on the trajectory, one may obtain a good description of these surfaces by considering sufficient points in $\delta A(\epsilon)$. In our calculation we have taken neighbourhoods of one hundred points, and verified that the qualitative behaviour of the surfaces does not change when taking two hundred points. In addition we have taken radii $\epsilon = 9 \delta x/10$ where $\delta x$ is the minimum distance between the points $A$ that constitute the parameter space (for a regular partition of the parameter space); this selection guarantees to detect any change of the fidelity and Bures distance, and also it does not associate the same value to two neighbouring points in the partition.

\begin{figure}
	\begin{center}
		{\includegraphics[width=0.6\linewidth]{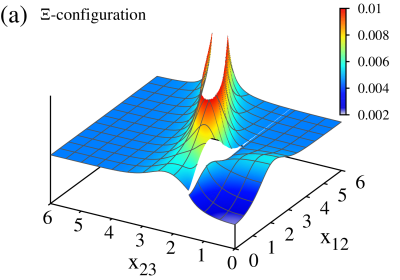}}\\
		{\includegraphics[width=0.6\linewidth]{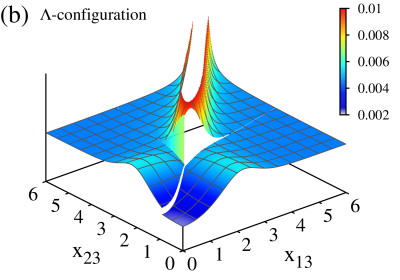}}\\
		{\includegraphics[width=0.6\linewidth]{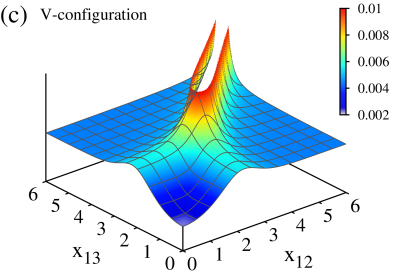}}
	\end{center}
	 \caption{(Colour online) Maximum Bures distance surfaces for the different atomic configuratios: $\Xi, \,\Lambda, \, V$.}
\label{maxburessurface}
\end{figure}

The shape of ${D_B}_{max}$ for the three atomic configurations is shown in Fig.~\ref{maxburessurface}. Note that the discontinuities and local maxima correspond perfectly with the separatrices in figure~\ref{quant_sep}. A change of parity in the ground state occurs precisely where these surfaces show discontinuities.

\section*{References}


\providecommand{\newblock}{}

\end{document}